\documentclass{easychair}

\usepackage{doc}

\title{Improved Polar-code-based Efficient Post-processing Algorithm for Quantum Key Distribution}

\author{
	Junbin~Fang\inst{1}\inst{3}\footnotemark[1]
\and
    Zhengzhong~Yi\inst{2}\footnotemark[1]
\and
    Jin~Li\inst{1}
\and 
	Zhipeng~Liang\inst{2}
\and 
	Yulin~Wu\inst{2}
\and 
	Wen~Lei\inst{1}
\and 
	Zoe~Lin~Jiang\inst{2}\footnotemark[2]
\and 
	Xuan~Wang\inst{2}\inst{4}\footnotemark[2]

\footnotetext[1]{Junbin Fang and Zhengzhong Yi are the joint first authors, they contribute equally to the paper.}
\footnotetext[2]{Corresponding authors}
}

\institute{
 	Department of Optoelectronic Engineering, Jinan University, Guangzhou, 510632, China.\\
\and
	School of Computer Science and Technology, Harbin Institute of Technology, Shenzhen. Shenzhen, 518055, China.\\
\and
	Cyberspace Security Research Center, Peng Cheng Laboratory, Shenzhen 518055, China.\\
\and
	Pengcheng Laboratory, Shenzhen, Guangdong 518055, China\\
	\email{wangxuan@cs.hitsz.edu.cn}, \email{zoeljiang@hit.edu.cn}
}

\authorrunning{Fang, Yi, Li, Liang, Wu, Lei, Jiang and Wang}

\titlerunning{Improved Efficient Post-processing Algorithm for QKD}

\begin{document}

\maketitle

\begin{abstract}
  Combined with one-time pad encryption scheme, quantum key distribution guarantees the unconditional security of communication in theory. However, error correction and privacy amplification in the post-processing phase of quantum key distribution result in high time delay, which limits the final secret key generation rate and the practicability of quantum key distribution systems. To alleviate this limitation, this paper proposes an efficient post-processing algorithm based on polar codes for quantum key distribution. In this algorithm, by analyzing the channel capacity of the main channel and the wiretap channel respectively under the Wyner’s wiretap channel model, we design a codeword structure of polar codes, so that the error correction and privacy amplification could be completed synchronously in a single step. Through combining error correction and privacy amplification into one single step, this efficient post-processing algorithm reduces complexity of the system and lower the post-processing delay. Besides, the reliable and secure communicaiton conditions for this algorithm has been given in this paper. Simulation results show that this post-processing algorithm satisfies the reliable and secure communication conditions well.
\end{abstract}



%
%

\section*{Keywords}
Quantum key distribution, post-processing, polar codes, error correction, privacy amplification
\section{Introduction}
\label{intro}
Combined with one-time pad encryption scheme, quantum key distribution (QKD) can guarantee the unconditional security of communication system in theory\cite{lo2001simple,gottesman2004security,koashi2009simple,molotkov2001simple,leverrier2009unconditional,lo2001proof,gottesman2003proof,lo1999unconditional,shor2000simple}.
Unlike the traditional encryption schemes such as RSA and Elliptical Curves whose security is based on the complexity of certain mathematical problems and hence will be influenced by the computing power of computing devices, QKD's security is based on physics law and the degree of the perfection of practical devices, which will not be influenced by computing power. Hence, in the post-quantum era during which most of the traditional encryption schemes are challenged with the formidable computing power of quantum computation, researchers have attached great attention to QKD. However, most practical QKD systems take photons as secret key carriers \cite{gisin2002quantum,bennett2014quantum}, which makes these systems susceptible to device defect and results in bit error and information leakage. Therefore, it’s necessary to perform error correction (also known as secret-key reconciliation) and privacy amplification in the post-processing phase to correct the error bit and eliminate the information leakage. Unfortunately, these two steps increase system overhead and introduce high time delay, which has become a bottleneck of realizing high-speed QKD and limits the further practicability of QKD systems \cite{diamanti2016practical,fung2010practical}. 
The earliest error correction algorithm for QKD post-processing is BBBSS algorithm\cite{bennett1992experimental} which iteratively applies dichotomic parity check. Based on BBBSS algorithm, Brassard and Salvail proposed the Cascade algorithm\cite{brassard1993secret} which improves the error correction efficiency of BBBSS. However, both of these two algorithms need repetitive exchange of the checking information between Alice (information sender) and Bob (information receiver) in the public channel, which leads to low error correction efficiency and high time delay in the post-processing phase. To reduce this repetitive information exchange in the public channel, Winnow algorithm\cite{buttler2003fast}, in which the checking information only needs transmitting for once, was proposed in 2003. However, within the security threshold of qubit error rate (QBER), Winnow still has low error correction efficiency. In 2004, Pearson proposed to apply LDPC codes in QKD post processing\cite{pearson2004high}. This idea has been followed by researchers for many years\cite{elkouss2009efficient,elkouss2010information,walenta2014fast,yuan201810,mao2019high}. Though LDPC codes do improve the efficiency of error correction, its parity-check matrix relies on QBER and hence the error correction performance is quite sensitive to QBER. To overcome this shortcoming, Elkouss, Martinez-Mateo and Martin\cite{elkouss2010information} proposed auto-adaptive LDPC for QKD system, but the iterative decoding of LDPC still results in high decoding overhead. In 2014, Joduget and Kunz-Jacques\cite{jouguet2012high} first applied polar codes, whose code rate has been proved to achieve Shannon limit, to QKD, and discussed the feasibility. Later research\cite{lee2018improved} shows under short code length, the efficiency of polar codes is higher than LDPC codes’. In the past five years, the application of polar codes in QKD system has drawn the attention of researchers \cite{renes2013efficient,yan2018improved,yi2019efficient,nakassis2017polar,kim2017reconciliation}.

In the aspect of privacy amplification, at present, a universal class of hash functions\cite{bennett1995generalized} was widely used in information compression to guarantee the security of secret key. However, due to its high computation complexity, this scheme has high time delay. To lower the time delay, researchers applies Toeplitz hashing, which becomes the most widely used privacy amplification method in recent years\cite{yuan201810,Hayashi2011Exponential,zhang2012real,LiHigh}. By combining Toeplitz hashing with fast Fourier transform, researchers has reduced the computation complexity of Toeplitz hashing to $O(n {\rm log} n)$.

To provide a new idea to reduce the complexity and lower the time delay of the post-processing phase in QKD systems, a polar-code-based efficient QKD post-processing algorithm is proposed in this paper. Using Wyner’s wiretap channel model, we design a codeword structure of polar codes which satisfies the reliability and security for QKD post-processing. This idea has been applied to different communication systems in recent years\cite{che2017physical,chen2018secret}. By doing this, the error correction and privacy amplification which are the most time-consuming steps in the QKD post-processing could be completed synchronously in a single encoding and decoding process. Therefore, the complexity and time delay of post-processing can be reduced, and the final key generation rate can be improved. This will help with breaking through the bottleneck of realizing high-speed QKD system and promote practicability of QKD.

In 2019, we proposed polar codes-based one-step post-processing for quantum key distribution in our previous work\cite{Lijing2019}. However, there are three main drawbacks in our previous work. First, the security condition (see equation (5) in \cite{Lijing2019}) is inaccurate and ambiguous. Thus we modify the security condition in this paper (see equation (5) in this paper). Second, the protocol in  \cite{Lijing2019} is incomplete which may result in decoding failure and insecurity (see the steps 1 to 10 and figure 3 in \cite{Lijing2019}). In this paper, we modify the protocol (see the steps 1 to 10 and figure 4 in this paper), which makes it more reliable and secure. The last but the most important point is that our previous work lacks experimental verification, since we only calculated the coding rate, and analyzed the reliability and security in theory. In this paper, we verify the reliability and security of the protocol through a large number of simulation experiments (see the whole section – Section IV “Simulation results”).

The rest of this paper is organized as follows. In Section II, we introduce the basic theory about Wyner’s wiretap channel model, the secrecy capacity of discrete variable QKD (DVQKD) systems and polar codes. Then in Section III, polar-code-based efficient QKD post-processing algorithm is introduced. Section IV gives the simulation experiment result on code rate, decoding reliability and communication security. In Section V, we summarize our work.

\section{Basic Theory}
\label{2}
\subsection{Wyner’s Wiretap Channel Model}
\label{2.1}
\begin{figure}
	\includegraphics[scale=0.43]{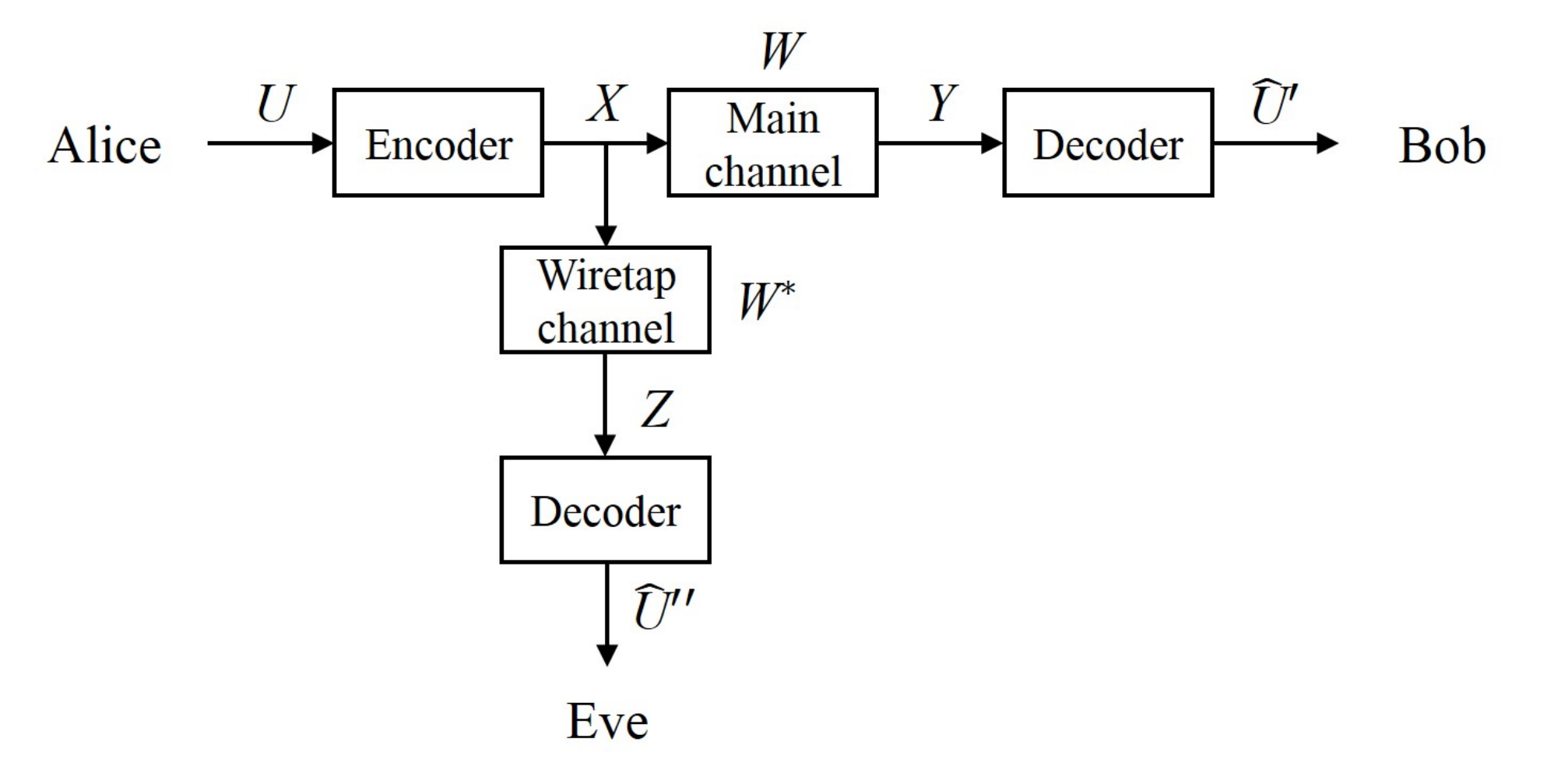}
	\caption{Wiretap channel model.}
	\label{fig:Wiretap channel model}
\end{figure}

The goal of secret communication is to realize reliable and secure information transmission between two authentic communication sides even under eavesdropping. The channel under eavesdropping can be depicted by Wyner’s wiretap channel model\cite{wyner1975wire} which is shown in Fig. \ref{fig:Wiretap channel model}. Authentic information sender Alice encodes the original information $U$ of length $k$ to code $X$ of length $n$ and sends code $X$ to authentic information receiver Bob through the main channel $W$, after which Bob gets information $Y$. In the meantime, eavesdropper Eve eavesdrops through the wiretap channel $W^*$ and gets information $Z$. After decoding, Bob gets the estimation ${\widehat U}'$ of original information $U$ and Eve gets the estimation $\widehat{U}''$.

In the Wyner’s wiretap channel model, when the wiretap channel $W^*$ is degenerative with respect to the main channel $W$ (that is to say, the channel capacity of the wiretap channel $C(W^*)$ is smaller than the channel capacity of the main channel $C(W)$), with the code length tending to infinite, one can design a secure coding scheme which satisfies the communication reliability and security. Furthermore, the largest code rate is equal to the secrecy capacity $C_{\rm sec}$ which is defined by $C_{\rm sec} \equiv C(W)-C(W^*)$. In other words, for all $\epsilon>0$, there exist coding schemes of rate $R \ge C_{\rm sec}-\epsilon$ that asymptotically achieve both the reliability and the security objectives\cite{mahdavifar2011}. Here, the reliability is measured by the decoding bit error rate (BER) of Bob, and the security is measured by the mutual information of $\widehat{U}''$ and $U$. Reliable communication means that

\begin{equation}
\lim_{n \to \infty} Pr(\widehat{U}'_i \neq U_i) = 0,
\label{eq:reliability1}
\end{equation}

where the subscript $i$ means the $i$th bit in $\widehat{U}'$ and $U$. Secure communication means that

\begin{equation}
\lim_{n \to \infty} I(\widehat{U}_i'';U_i) = 0,
\label{eq:security1}
\end{equation}

where  $I(\widehat{U}_i'';U_i)$ is the mutual information between Alice and Eve, $\widehat{U}_i''$ is the $i$th bit in $\widehat{U}''$. Combining equation (\ref{eq:security1}) with the relation between mutual information $I(\widehat{U}_i'';U_i)$ and conditional entropy $H(U_i|\widehat{U}_i'')$ depicted by equation (\ref{eq:mutualinformation}), and the definition of conditional entropy depicted by equation(\ref{eq:informationentropy}),

\begin{equation}
I(\widehat{U}_i'';U_i)\equiv H(U_i)-H(U_i|\widehat{U}_i'')=1-H(U_i|\widehat{U}_i''),
\label{eq:mutualinformation}
\end{equation}

\begin{equation}
H(U_i|\widehat{U}_i'') \equiv -\sum_{a \in U_i}\sum_{b \in \widehat{U}_i''}p(a,b){\rm log}p(a|b),
\label{eq:informationentropy}
\end{equation}

we can rewrite eqaution (\ref{eq:security1}) to 

\begin{equation}
\lim_{n \to \infty}Pr(\widehat{U}_i'' \neq U_i)=\lim_{n \to \infty}Pr(\widehat{U}_i'' = U_i)=0.5.
\label{eq:security2}
\end{equation}

Equation (\ref{eq:reliability1}) is the \emph{reliable communication condition} and equation (\ref{eq:security2}) is the \emph{secure communication condition}. They imply that a reliable and secure coding scheme demands that, with code length tending to infinite, Bob asymptotically achieves 0 and the decoding BER of Eve asymptotically achieves 0.5.  

\subsection{Channel Capacity of DVQKD Systems}
\label{2.2}
In QKD systems, after qubit transmission and sifting, Alice obtains sifted key $KA_{\rm sifted}$ and Bob obtains sifted key $KB_{\rm sifted}$. Due to the defect of devices, channel noise and possible eavesdropping in the practical QKD system, in general, $KA_{\rm sifted} \neq KB_{\rm sifted}$. Namely, there are error bits. Denote the bit error rate in practical QKD system by $p$.

DVQKD is the maturest and the most widely used QKD system. For those DVQKD systems which apply BB84 protocol, their qubit transmission channel can be regarded as binary symmetric channel (BSC). Under this assumption, the mutual information between Alice and Bob is 
\begin{equation}
I_{\rm AB}=1-h_2(p),
\label{eq:IAB}
\end{equation}
where $h_2(\cdot)$ is binary entropy function\cite{elkouss2009efficient}. Considering the maximum safety of communication, we can regard all the noise in practical systems results from eavesdropping. Hence, all information Eve can obtain is at most 
\begin{equation}
I_{\rm AE}=h_2(p).
\label{eq:IAE}
\end{equation} 
If we adopt Wyner’s wiretap channel model to depict QKD system, the channel capacity of main channel $W$ is 
\begin{equation}
C(W)=I_{AB}=1-h_2(p),
\label{eq:CW}
\end{equation}
the channel capacity of the wiretap channel is 
\begin{equation}
C(W^*)=I_{\rm AE}=h_2(p),
\label{eq:CW*1}
\end{equation}
and the secrecy capacity is 
\begin{equation}
C_{\rm sec}=C(W)-C(W^*)=1-2h_2(p).
\end{equation}
The secrecy capacity is equal to the secure final key generation rate $k_{\rm th}$\cite{gottesman2004security}.

Practical DVQKD systems require that $k_{\rm th}=1-2h_2(p) \ge 0$. This means that the value range of QBER p is [0, 0.11] and $C(W) \ge C(W^*)$. Hence, according to the Wyner’s wiretap channel model theory mentioned in Section II. A, in this situation, we can design a coding scheme which achieves the secrecy capacity.

\subsection{Polar Codes}
\label{2.3}
Polar codes are the only coding scheme which has been proved in theory that their code rate can achieve Shannon limit\cite{arikan2009channel}. Besides, the encoding and decoding complexity of polar codes is relatively small compared with LDPC codes\cite{arikan2009channel}. Through recursively polarizing $N$ independent identically distributed (i.i.d.) channels whose capacity are all $C$, one can get $N$ coordinate subchannels whose capacity polarizes – with the growth of code length $N$, the capacity of $N \cdot C$ coordinate subchannels asymptotically tends to 1, while the capacity of the other $N \cdot (1-C)$ coordinate subchannels asymptotically tends to 0. That is to say, the former $N \cdot C$ coordinate subchannels are optimized and the latter $N \cdot (1-C)$ coordinate subchannels are degraded. The optimized channels will be used to transmit information bits and the degraded ones will be used to transmit frozen bits. Hence, the code rate asymptotically achieves the channel capacity which equals to $N \cdot C$.

\begin{figure}
	\includegraphics[scale=0.35]{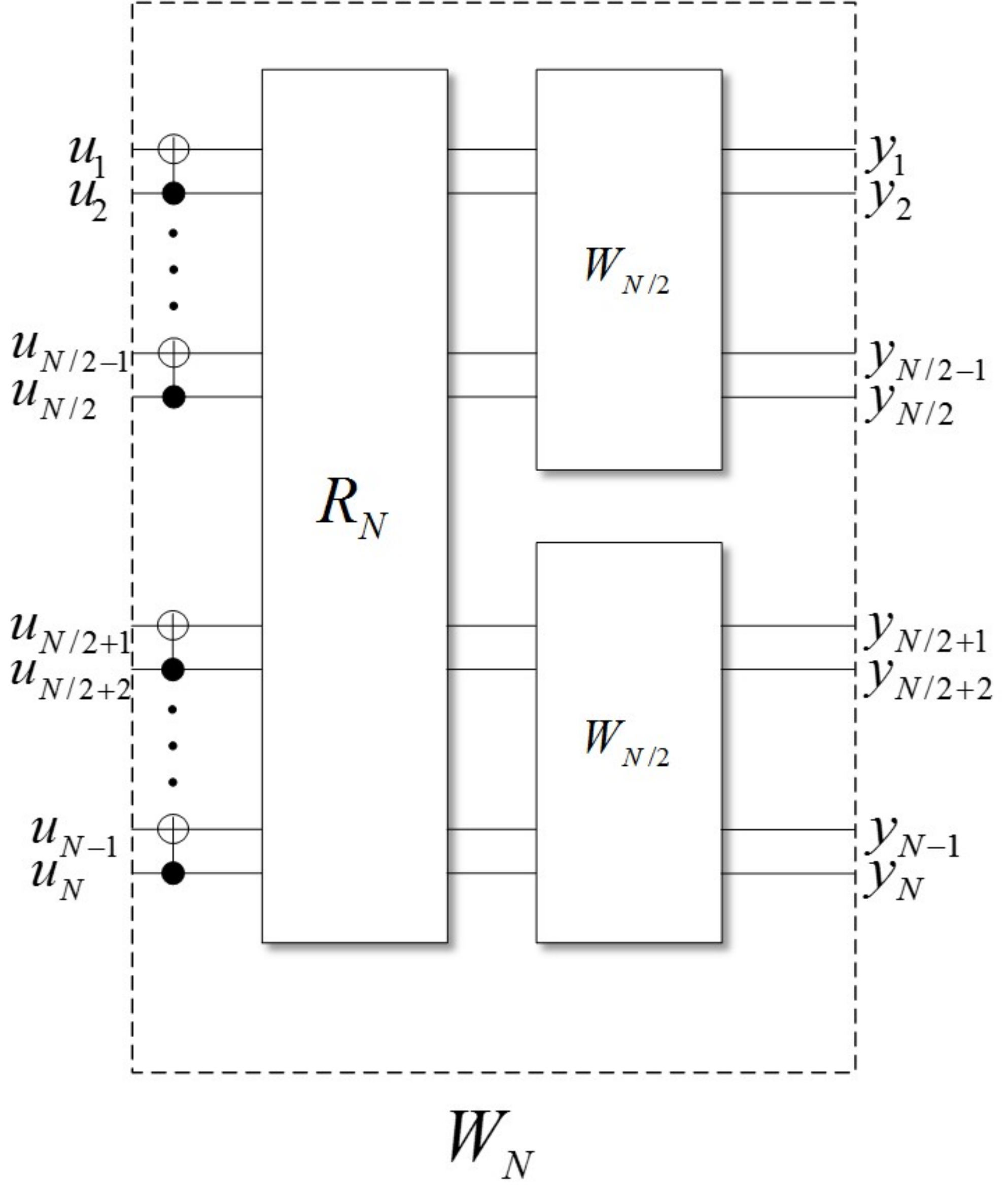}
	\caption{Channel polarization. $R_N$ is the bit-reversal operation. When $N=1$, $W_1$ is the original channel $W$.}
	\label{fig:Channel polarization}
\end{figure}

Denote the original $N$ i.i.d. channels by $W$. As shown in Fig. \ref{fig:Channel polarization}, through channel combining in a recursive way, we get the combining channel $W_N$ of all $N$ i.i.d. channels. Then through channel splitting, we can obtain $N$ coordinate subchannels  $W_N^{(i)}$\cite{arikan2009channel}. The superscript $(i)$ means the $i{\rm th}$ subchannel. In the rest of this paper, $1\leq i \leq N$.

Under finite code length $N$, we need to evaluate the channel quality of each coordinate subchannel. According to the channel quality, we rank all coordinate subchannels in descending order. Then, the first $K$ of them are chosen to transmit information bits according to concrete error correction requirement. In this way, the construction of polar codes is fulfilled. It’s noticeable that the determination of $K$ will impact the reliability and the code rate of the code structure we design – if $K$ is too high, the decoding reliability will be unacceptable; if it is too low, the channel-capacity-reachable characteristic of polar code cannot be fully used and hence the code rate will be unsatisfactory. $K$ can be determined by setting target frame error rate (TFER, it is a predefined value which Alice and Bob try to make the practical frame error rate of their communication lower than through error correction), which is used in our algorithm in Section III.

At present, there are several ways to realize the construction of polar codes\cite{arikan2009channel,tal2013construct,mori2010properties,mori2009performance}. In this paper, we adopt Tal’s method\cite{tal2013construct} to construct polar codes, in which the probability of error $P_{\rm e}(W_N^{(i)})$ under maximum-likelihood decision is used to measure the quality of each coordinate subchannel $W_N^{(i)}$. Through a asymptotic method called channel degradation\cite{tal2013construct}, we calculate the upper bound of each $P_{\rm e}(W_N^{(i)})$ which will be used to construct polar codes.

\section{Polar-code-based Efficient Post-processing Algorithm}
\label{3}
Error correction and privacy amplification are two crucial steps in QKD post-processing. The goal of error correction is to eliminate the difference between Alice’s sifted key $KA_{\rm sifted}$ and Bob’s sifted key $KB_{\rm sifted}$ through information exchange between Alice and Bob, so that they can obtain the information which is equal to the capacity $C(W^*)$ of the main channel. The goal of privacy amplification is to compress the exchanged information between Alice and Bob to remove the information Eve can obtain, which is equal to the capacity $C(W^*)$ of wiretap channel.

Aiming at these two functions of the two crucial steps, we propose an efficient post-processing algorithm which can fulfill error correction and privacy amplification at the same time. This algorithm is called polar-code-based efficient post-processing (PCEP) algorithm. The concrete steps of PCEP are as follows. Denote the TFER by $FER_{\rm target}$, the target privacy amplification index (TPAI, it is a predefined value which Alice and Bob try to make the practical privacy amplification index lower than. Privacy amplification index is the leaked information rate, which is equal to the amount of leaked information leaked in a single code block divided by the code block length) by $PAI_{\rm target}$. 

\textbf {Step 1}: Parameter estimation

Alice and Bob compare the bases they use in the qubit transmission phase and get their own sifted key $KA_{\rm sifted}$ and $KB_{\rm sifted}$. Then they choose some bits from their own sifted key to estimate the bit error rate $p_{\rm m}$ (to distinguish the indexes of main channel and wiretap channel, we write an “m” in the subscript to represent that this index belongs to “main channel” or a “w” to represent that this index belongs to “wiretap channel”) in the main channel as in other common post-processing algorithm . If $p_{\rm m}$ exceeds the security threshold, they abort this key distribution, or else they enter into next step.

\textbf{Step 2}: Polarization of the main channel

Alice and Bob polarize the main channel $W$ by Arikan’s method[36] and obtain $N$ coordinate subchannels $W_N^{(i)}$.

\textbf{Step 3}: Channel quality evaluation in the main channel

Denote the code length that Alice and Bob use by $N$, Alice and Bob take $p_{\rm m}$ as the channel quality index of the main channel, according to which they adopt Tal’s polar code construction algorithm\cite{tal2013construct} to calculate the upper bound $UP_{\rm e,m}(W_N^{(i)})$ (not necessarily the supremum) of the decoding error rate $P_{\rm e,m}(W_N^{(i)})$ under maximum-likelihood decision of each coordinate subchannel $W_N^{(i)}$. $UP_{\rm e,m}(W_N^{(i)})$ are used to evaluate the channel quality of each coordinate subchannel, the lower the better.

\textbf{Step 4}: Optimized coordinate subchannels selection in the main channel

Alice and Bob sort all coordinate subchannels $W_N^{(i)}$ according to $UP_{\rm e,m}(W_N^{(i)})$ \emph{in ascending order}, and chooses the first $K_{\rm m}$ coordinate subchannels which satisfy Equation (\ref{eq:GNpick}) to compose the optimized channel set $G_{N}(W, FER_{\rm target})$. The rest of coordinate subchannels compose the degraded channel set $B_{N}(W, FER_{\rm target})$.

\begin{equation}
\sum_{i}UP_{\rm e,m}(W_N^{(i)}) \leq FER_{\rm target}.
\label{eq:GNpick}
\end{equation}

That is to say, Alice and Bob divide all coordinate subchannels in the main channel to two sets:
\begin{equation}
G_{N}(W,FER_{\rm target})\equiv\{i|1\leq i \leq N\} \cap \{i|\sum_{i}UP_{\rm e,m}(W_N^{(i)})\leq FER_{\rm target}\},
\label{eq:GN}
\end{equation}

\begin{equation}
B_{N}(W,FER_{\rm target})\equiv\{i|1\leq i \leq N\}\setminus G_{N}(W,FER_{\rm target}).
\label{eq:BN}
\end{equation}

From equation (\ref{eq:GN}) and (\ref{eq:BN}), we can see that $G_{N}$ and $B_{N}$ are functions of $W$ and $FER_{\rm target}$. This is why we write $G_{N}$ as $G_{N}(W,FER_{\rm target})$ and $B_{N}$ as $B_{N}(W,FER_{\rm target})$. For convenience, $G_{N}$ and $B_{N}$ will be used in the rest of this paper.

\textbf{Step 5}: Polarization of the wiretap channel

Alice and Bob polarize the wiretap channel $W^*$ by Arikan’s method\cite{arikan2009channel} and obtain $N$ coordinate subchannels $W_N^{(i)}$.

\textbf{Step 6}: Channel quality evaluation in the wiretap channel

Alice and Bob calculate the bit error rate $p_{\rm w}$ of wiretap channel according to $I_{\rm AE}=1-h_2(p_{\rm w}) = h_2(p_{\rm m})$ as mentioned in Section II. B. Then they take $p_{\rm w}$ as the channel quality index of the wiretap channel, according to which they adopt Tal’s polar codes construction algorithm\cite{tal2013construct} to calculate the upper bound  $UP_{\rm e,w}(W_N^{*(i)})$ (not necessarily the supremum) of the probability of error $P_{\rm e,w}(W_N^{*(i)})$ under maximum-likelihood decision of each coordinate subchannel $W_N^{*(i)}$ in wiretap channel. Using equation (\ref{eq:CW*2}), Alice and Bob calculate the channel capacity of $C_{\rm w}(W_N^{*(i)})$ each coordinate subchannel.

\begin{equation}
C_{\rm w}(W_N^{*(i)})=1-h_2(P_{\rm e,w}(W_N^{*(i)})).
\label{eq:CW*2}
\end{equation}

The channel capacity $C_{\rm w}(W_N^{*(i)})$ is used to evaluate the channel quality of each coordinate subchannel, the higher the better.

\textbf{Step 7}: Optimized coordinate subchannels selection in the wiretap channel

Alice and Bob sort all coordinate subchannels $W_N^{*(i)}$ according to $C_{\rm w}(W_N^{*(i)})$ \emph{in ascending order} and chooses the first $K_{\rm w}$ ones which satisfy equation (\ref{eq:BN*pick}) to compose degraded channel set $B_{N}^*(W^*, PAI_{\rm target})$ with respect to Eve. The rest of coordinate subchannels compose optimized channel set $G_{N}^*(W^*, PAI_{\rm target})$ with respect to Eve.

\begin{equation}
\sum_i C_{\rm w}(W_N^{*(i)}) \leq PAI_{\rm target}.
\label{eq:BN*pick}
\end{equation}

That is to say, Alice and Bob divide all coordinate subchannels in the wiretap channel to two sets:

\begin{equation}
B_{N}^*(W^*,PAI_{\rm target})\equiv\{i|1\leq i \leq N\} \cap \{i|\sum_{i}C_{\rm w}(W_N^{*(i)})\leq PAI_{\rm target}\},
\label{eq:BN*}
\end{equation}

\begin{equation}
G_{N}^*(W^*,PAI_{\rm target})\equiv\{i|1\leq i \leq N\}\setminus B_{N}^*(W^*,PAI_{\rm target}).
\label{eq:GN*}
\end{equation}

From equation (\ref{eq:BN*}) and (\ref{eq:GN*}), we can see that $B_{N}^*$ and $G_{N}^*$ are functions of $W^*$ and $PAI_{\rm target}$. This is why we write $G_{N}^*$ as $G_{N}^*(W^*,PAI_{\rm target})$ and $B_{N}^*$ as $B_{N}^*(W^*,PAI_{\rm target})$. For convenience, $G_{N}^*$ and $B_{N}^*$ will be used in the rest of this paper.

\textbf{Step 8}: Determination of code structure

\begin{figure}
	\includegraphics[scale=0.35]{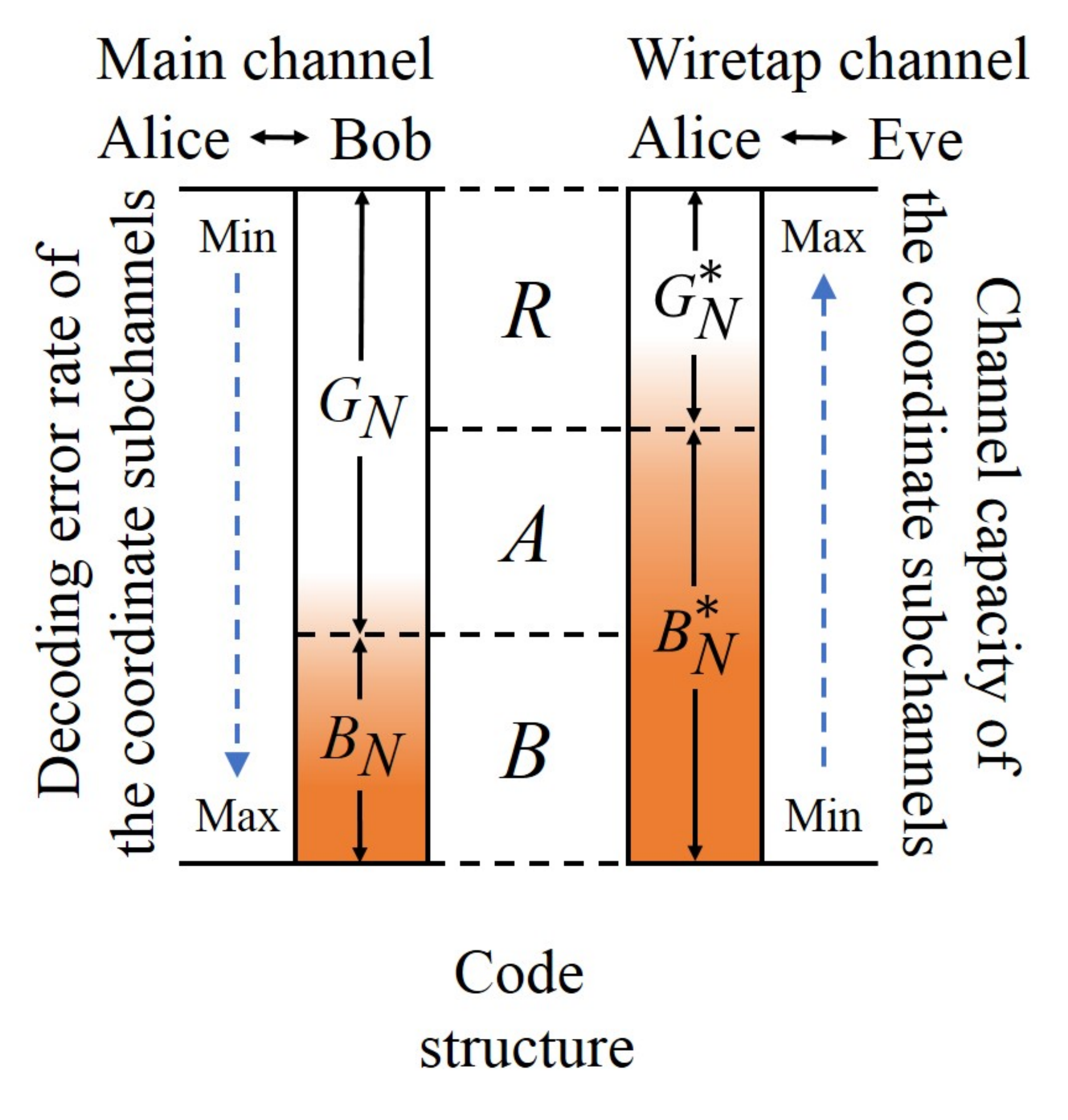}
	\caption{Code construction. The two columns which are colored by gradient represent the coordinate subchannels of the main channel and the wiretap channel. The deeper the color is, the worse the channel quality is.}
	\label{fig:code construction}
\end{figure}

After the above steps, Alice and Bob obtain four sets of coordinate subchannels. The first set $G_N$ is the optimized coordinate subchannels to Bob, the second set $B_N$ is the degraded ones to Bob, the third set $G_N^*$ is the optimized ones to Eve, and the last set $B_N^*$ is the degraded ones to Eve. As shown in Fig. \ref{fig:code construction}, the subchannels which belong to $B_N$ must belong to $B_N^*$, and the ones which belong to $G_N^*$ must belong to $G_N$. This is because that the wiretap channel is degenerative with respect to the main channel. Therefore, those subchannels which are degraded to Bob must be degraded to Eve, and those which are optimized to Eve must be optimized to Bob. Hence, $G_N$ and $B_N^*$ have intersection.

Based on the above analysis of the four sets $G_N$, $B_N$, $G_N^*$, and $B_N^*$, Alice and Bob can redivide all subchannels into three sets without intersection as follows.
\begin{equation}
R\equiv G_N^*,
\label{eq:R}
\end{equation}
\begin{equation}
A\equiv B_N^*\cap G_N,
\label{eq:A}
\end{equation}
\begin{equation}
B\equiv B_N.
\label{eq:B}
\end{equation}
Alice and Bob choose the subchannels in $A$ to transmit the information bits (in this situation, they are the bits of secret key), the subchannels in $R$ to transmit random bits, and the subchannels in $B$ to transmit frozen bits. By this redivision, the code structure is determined

\textbf{Step 9}: Code transmission

\begin{figure}
	\includegraphics[scale=0.36]{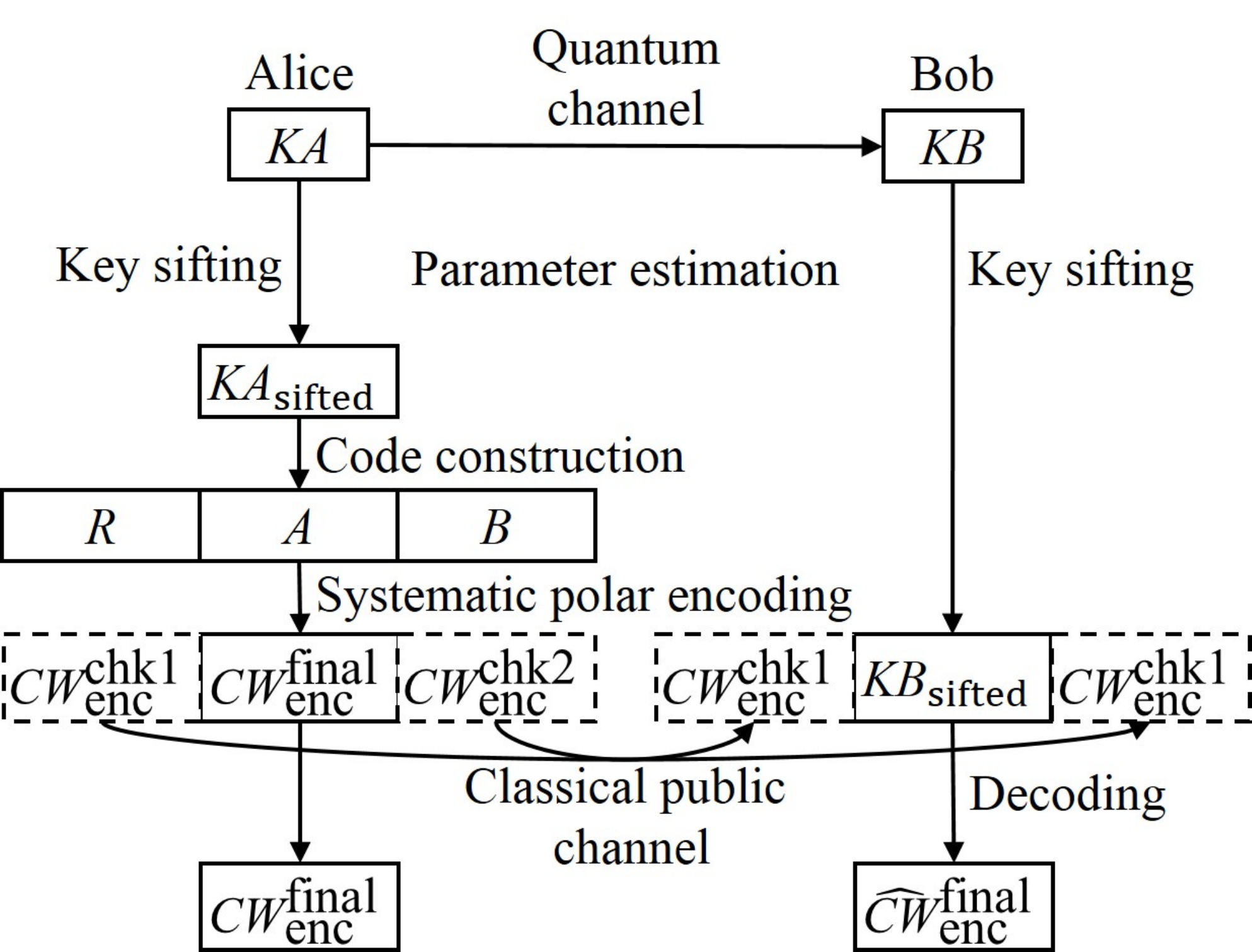}
	\caption{Polar-code-based efficient post-processing algorithm}
	\label{fig:PCEPalgorithm}
\end{figure}

Alice randomly generates the bits which belong to $R$, sets all bits which belong to $B$ to zero, and puts $KA_{\rm sifted}$ into the bits which belong to $A$. Then she connects them according to the order of corresponding coordinate subchannels to form the original code. After encoding the original code by systematic polar coding algorithm\cite{arikan2011systematic}, Alice gets code $CW_{\rm enc}$. As shown in Fig. \ref{fig:PCEPalgorithm}, $CW_{\rm enc}$ is composed of $CW_{\rm enc}^{\rm chk1}$, $CW_{\rm enc}^{\rm final}$(under systematic polar coding,  $CW_{\rm enc}^{\rm final}=KA_{\rm sifted}$) and $CW_{\rm enc}^{\rm chk2}$, which are the systematic polar encoding results of the bits belong to $R$, $A$, and $B$, respectively. Alice only sends the check bits $CW_{\rm enc}^{\rm chk1}$ and $CW_{\rm enc}^{\rm chk2}$ to Bob through classical public channel.

\textbf{Step 10}: Error correction

Bob puts his sifted key $KB_{\rm sifted}$ into the bits which belong to $A$, puts $CW_{\rm enc}^{\rm chk1}$ into the bits which belong to $R$, and puts $CW_{\rm enc}^{\rm chk2}$ into the bits which belong to $B$. Then he decodes this bit string to get $\widehat{CW}_{\rm enc}^{\rm final}$. At last, Alice and Bob take $CW_{\rm enc}^{\rm final}$ and $\widehat{CW}_{\rm enc}^{\rm final}$ as their final key respectively. The reliable communication condition eqaution (\ref{eq:reliability2}) asks that
\begin{equation}
\lim_{n \to \infty} Pr(\widehat{CW}_{\rm enc}^{\rm final} \neq CW_{\rm enc}^{\rm final}) = 0.
\label{eq:reliability2}
\end{equation}

\section{Reliability And Security Analysis of PCEP Algorithm}
\label{4}
In PCEP algorithm, Bob gets $CW_{\rm enc}^{\rm chk1}$, $CW_{\rm enc}^{\rm chk2}$ and $KB_{\rm sifted}$ which is obtained through the quantum channel with bit error rate $p_{\rm m}$. Assume Eve has full access to the classical channel, and all that she can get is $CW_{\rm enc}^{\rm chk1}$, $CW_{\rm enc}^{\rm chk2}$ and $KB_{\rm sifted}$ which is obtained by eavesdropping the quantum channel with bit error rate $p_{\rm w}$. According to equation (\ref{eq:IAE}), we obtain
\begin{equation}
1-h_2(p_{\rm w}) =h_2(p_m).
\end{equation}

When wiretap channel $W^*$ is degenerative to main channel $W$, $p_{\rm m} < p_{\rm w}$. 
The key $KA{\rm sifted}$ has been encoded into $CW_{\rm enc}^{\rm final}$, and under systematic polar coding, $CW_{\rm enc}^{\rm final}=KA_{\rm sifted}$. To obtain the key, Bob decodes $CW_{\rm enc}^{\rm chk1}$, $CW_{\rm enc}^{\rm chk2}$ and $KB_{\rm sifted}$ to get $\widehat{CW}_{\rm enc}^{\rm final}$, Eve decodes $CW_{\rm enc}^{\rm chk1}$, $CW_{\rm enc}^{\rm chk2}$ and $KE_{\rm sifted}$ to get $\widehat{CW}_{\rm enc}^{\prime\rm final}$. Because the coordinate subchannels in set A is optimized to Bob but degraded to Eve, the code structure which is determined in step 8 is optimized to Bob but degraded to Eve. Hence, with the growth of code length $N$, the decoding error rate of Bob tends to 0 while the decoding error rate of Eve tends to 0.5 (namely, the information in the wiretap channel has been compressed to zero). That is to say, $\lim_{n \to \infty} Pr(\widehat{CW}_{\rm enc}^{\rm final} \neq CW_{\rm enc}^{\rm final}) = 0$ and $\lim_{n \to \infty} Pr(\widehat{CW}_{\rm enc}^{\rm final} = CW_{\rm enc}^{\prime \rm final}) = 0$, which satisfies the reliable communication condition equation (\ref{eq:reliability1}) and secure communication condition equation (\ref{eq:security2}). 

\section{Simulation Results}
\label{5}
To prove the feasibility of PCEP algorithm, we conduct a series of simulation experiment on code rate, reliability and security. It should be noticed that the range of $p_{\rm m}$ has been limited to $[0, 0.11]$ because as mentioned in Section II. B, only in this range is $W^*$ degenerative to $W$. In all simulation experiment, we set $FER_{\rm target}$ to 0.1 and $PAI_{\rm target}$ to $10^{-7}$.

\subsection{Code Rate}
\label{subsection_Code_Rate}

\begin{figure}
	\includegraphics[scale=0.43]{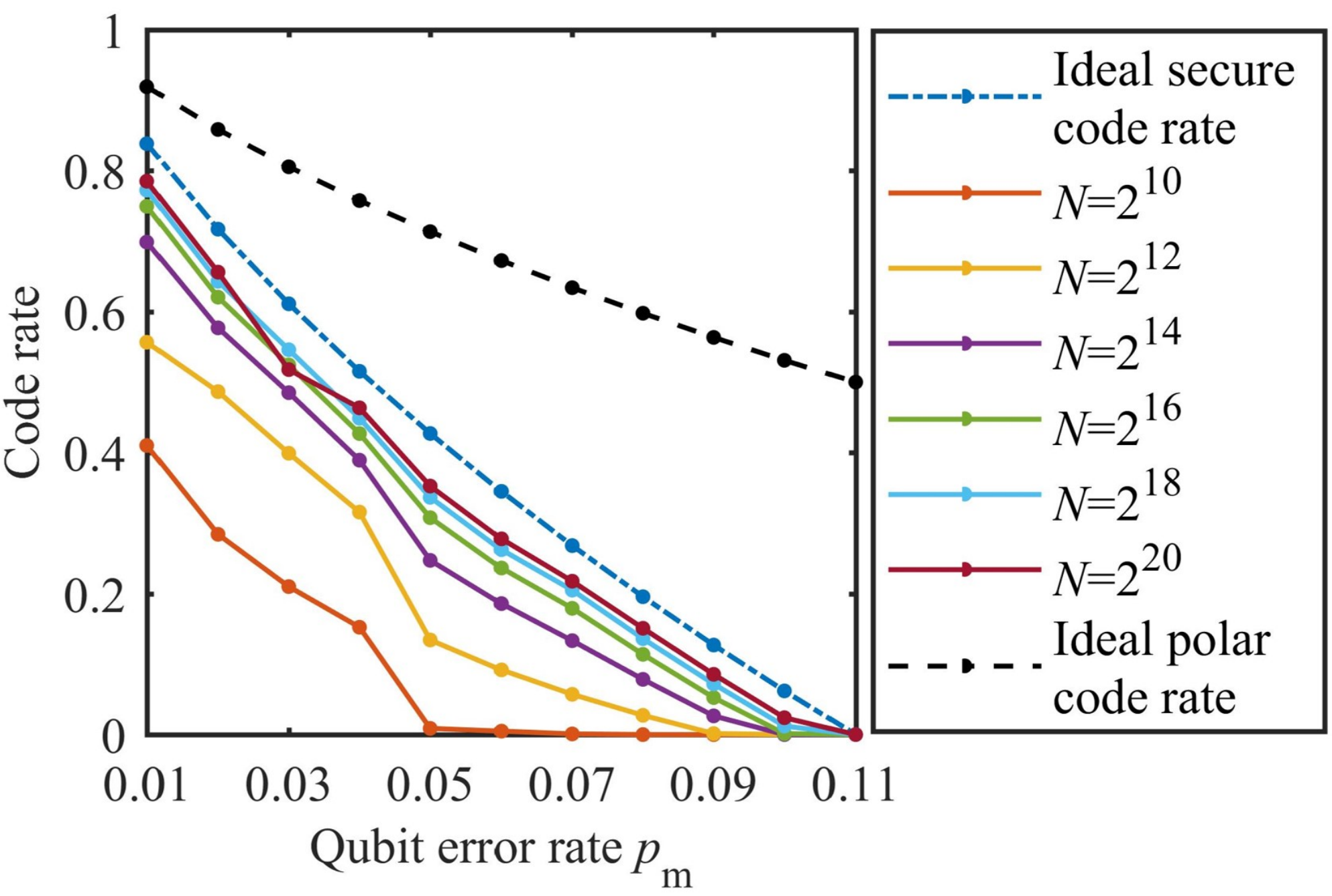}
	\caption{Code rate. The ideal polar code rate is the ideal code rate of polar code itself without considering wiretap channel. The ideal secure code rate is the ideal code rate of PCEP algorithm under wiretap channel model.}
	\label{fig:coderate}
\end{figure}

As shown in Fig. \ref{fig:coderate}, under different code length $N$, we calculate the code rate. It is observed that with the increase of QBER $p_{\rm m}$ of the main channel, the code rate tends to zero. Moreover, except a single point (where $N=2^{20}$, $p_{\rm m}=0.03$), under the same QBER $p_{\rm m}$, the longer the code length is, the higher the code rate is. This is in accord with the asymptotic property of polar codes.

\begin{figure}
	\includegraphics[scale=1]{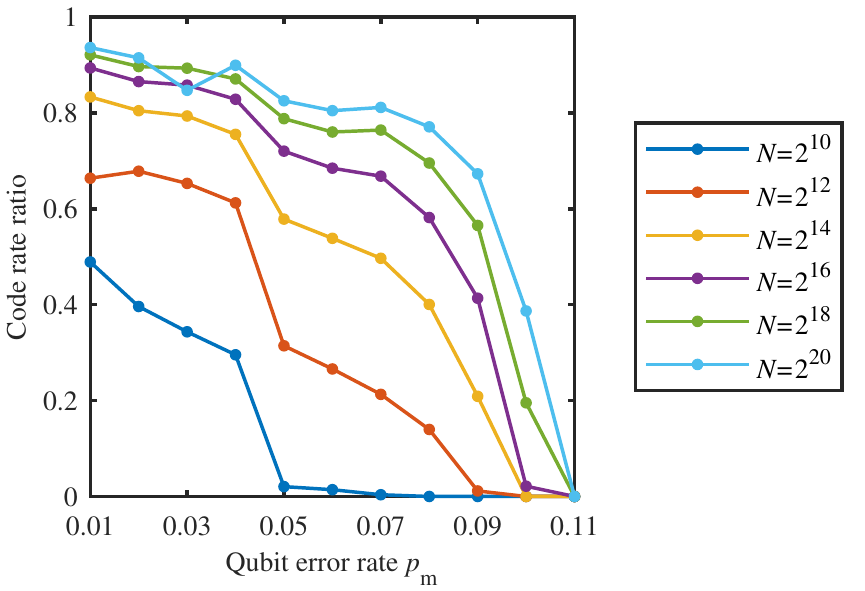}
	\caption{The ratio of the practical code rate and the theoretical secure code rate}
	\label{fig:coderateratio2}
\end{figure}

Fig. \ref{fig:coderateratio2} shows the ratio of the practical code rate and the theoretical secure code rate. It can be observed that with the increase of QBER $p_{\rm m}$, the ratio decreases to zero. The theoretical secure code rate can be regarded as a measurement of the error correcting capability of polar codes, while the practical code rate can be regarded as a measurement of the specific requirement for error correcting capability in certain setting. Therefore, the ratio can be used to measure the extent to which the requirement can be met – the lower the ratio is, the higher the extent is, and hence the better the error correcting performance is. Hence, the lower the ratio is, the higher the decoding reliability should be, which is consistent with the simulation result in Section \ref{subsection_reliability}.

\subsection{Security – The Decoding FER And BER of Eve}
\label{key}

\begin{figure}
	\includegraphics[scale=1]{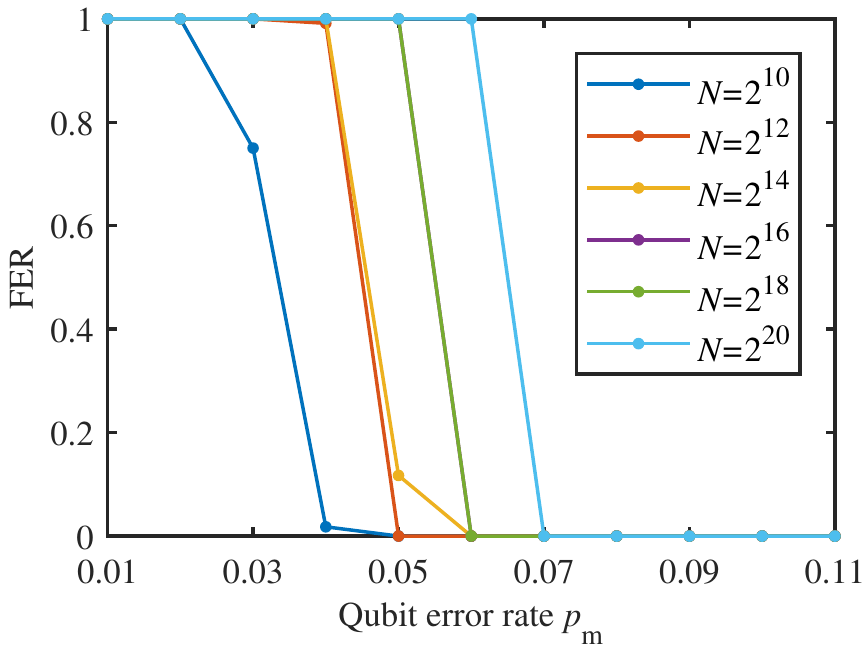}
	\caption{The decoding FER of Eve. The number of simulation tests is $1\times10^5$}
	\label{fig:FERofEve}
\end{figure}

\begin{figure}
	\includegraphics[scale=1]{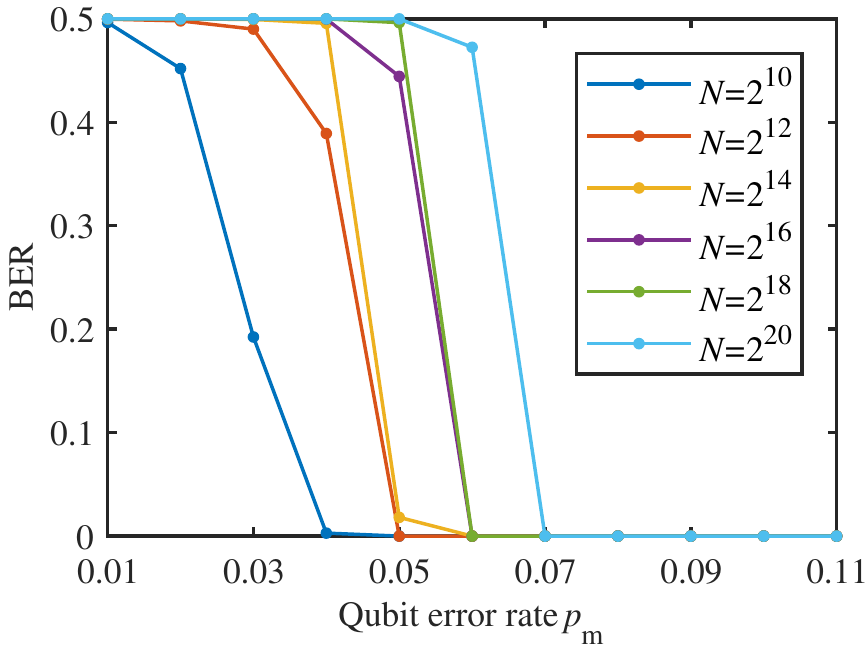}
	\caption{The decoding BER of Eve. The number of simulation tests is $1\times10^5$.}
	\label{fig:BERofEve}
\end{figure}

According to equation (\ref{eq:security2}), the security of PCEP algorithm can be measured by the decoding FER and BER of Eve, which is shown in Fig. \ref{fig:FERofEve} and Fig. \ref{fig:BERofEve}. It can be observed that when QBER $p_{\rm m}$ is small, the decoding FER and BER of Eve well satisfies the security condition equation (\ref{eq:security2}) ($FER=1$, $BER\sim 0.5$), while there is a threshold of QBER beyond which the decoding FER and BER of Eve dramatically decrease to zero. Moreover, the longer the code length, the higher the threshold, which coheres with the asymptotic property of polar codes.

\subsection{Reliability}
\label{subsection_reliability}
\begin{figure}
	\includegraphics[scale=0.44]{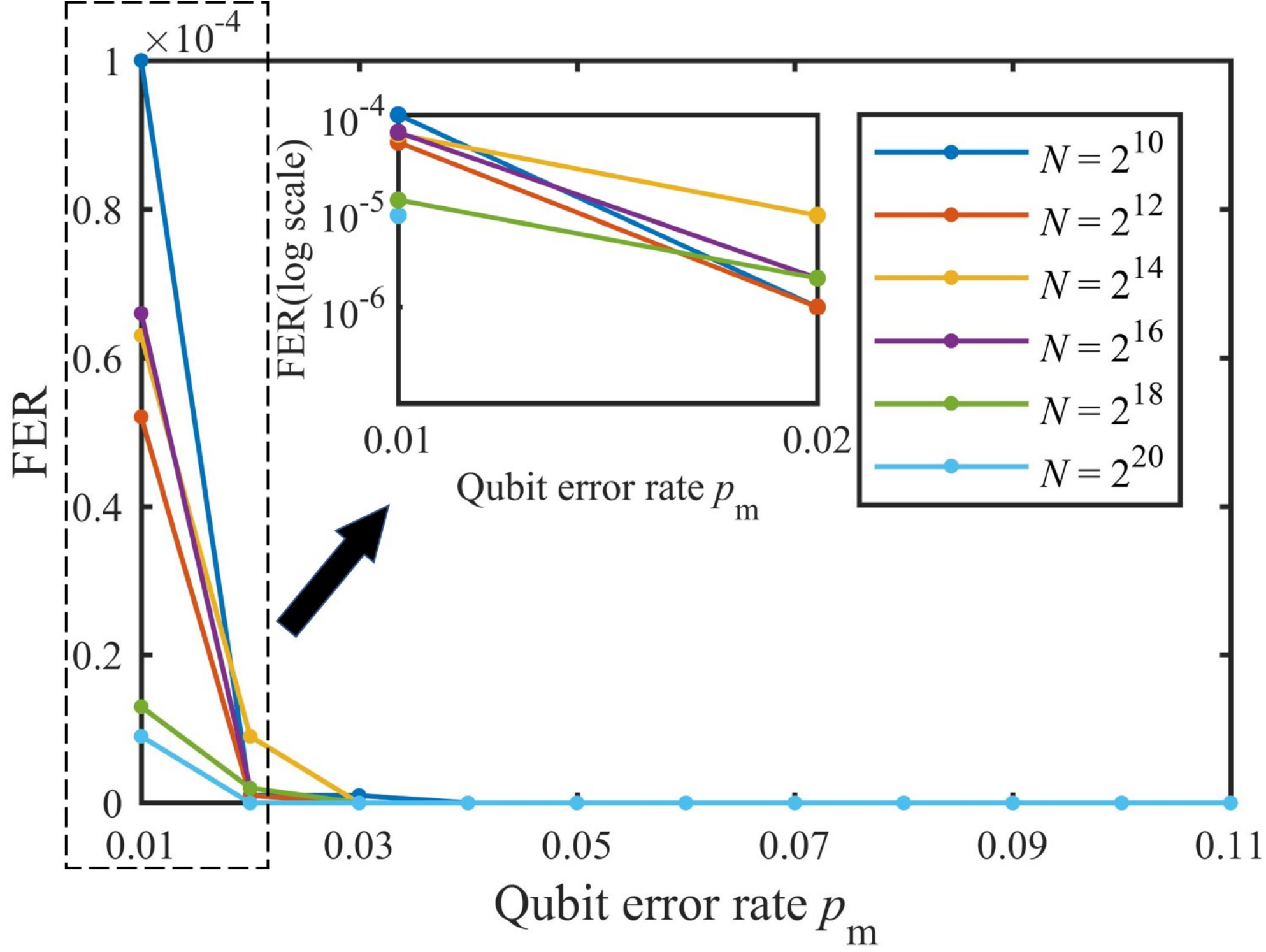}
	\caption{The decoding FER of Bob}
	\label{fig:FERofBob}
\end{figure}

\begin{figure}
	\includegraphics[scale=0.44]{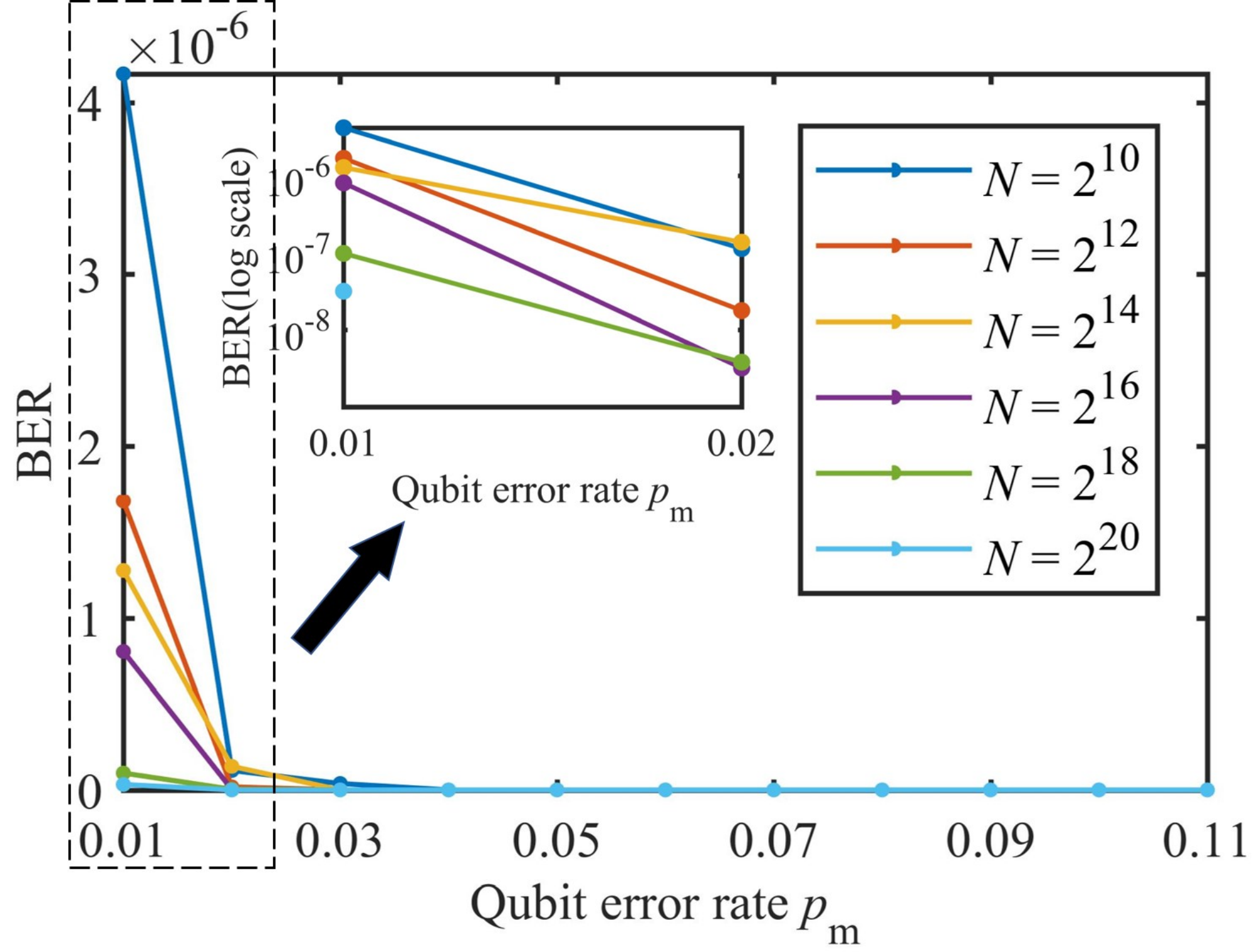}
	\caption{The decoding BER of Bob}
	\label{fig:BERofBob}
\end{figure}

According to equation (\ref{eq:reliability1}), the reliability of PCEP algorithm can be measured by the decoding FER and BER of Bob, which is shown in Fig. \ref{fig:FERofBob} and Fig. \ref{fig:BERofBob}. It is observed that the practical decoding FER and BER are satisfying under all code lengths shown in Fig. \ref{fig:FERofBob} and Fig. \ref{fig:BERofBob}. Besides, as shown in Fig. \ref{fig:FERofBob}, the maximum FER in the simulation is around $1\time10^{-4}$ when $N=2^{10}$ and $p=0.01$. Notice that the TFER has been set to 0.1, hence this target is well achieved. 

Moreover, under different code lengths, the decoding FER and BER of Bob decrease to zero rapidly with the increase of QBER $p_{\rm m}$. The reason for this counterintuitive phenomenon has been explained in the last paragraph in Section \ref{subsection_Code_Rate}.

\section{Conclusion}
\label{6}
In this paper, an efficient QKD post-processing algorithm PCEP which is based on polar codes is proposed. In PCEP algorithm, by analyzing the channel capacity of the main channel and the wiretap channel respectively under the Wyner’s wiretap channel model, we design a codeword structure of polar codes, so that the error correction and privacy amplification could be completed synchronously in a single encoding and decoding process. That is to say, PCEP algorithm realizes combining these two post-processing steps into one step. Through this, PCEP algorithm can reduce the complexity and lower the post-processing delay of QKD systems. This provides a new way to develop high-speed QKD systems. To clarify the reliability and security of PCEP algorithm, the reliability and security conditions have deen deduced from the perspective of information theory. Simulation results show that PCEP algorithm well satisfies the reliable and secure communication conditions. 
\section{Acknowledgments}

This work was supported in part by the National Key Research and Development Program of China (2018YFB1801900), in part by the National Natural Science Foundation of China (61771222, 61872109), in part by the Science and Technology Project of Guangzhou (201803020023), in part by the Project of Guangzhou Industry Leading Talents (CXLJTD-201607), in part by the Science and Technology Project of Shenzhen (JCYJ20170815145900474, JSGG20170824163239586), in part by the Peng Cheng Laboratory Project of Guangdong Province (PCL2018KP004), and in part by the Fundamental Research Funds for the Central Universities (21620439).

\label{sect:bib}
\bibliographystyle{plain}
\bibliography{mycitation}

\begin{thebibliography}{10}

\bibitem{arikan2009channel}
Erdal Arikan.
\newblock Channel polarization: A method for constructing capacity-achieving
  codes for symmetric binary-input memoryless channels.
\newblock {\em IEEE Transactions on information Theory}, 55(7):3051--3073,
  2009.

\bibitem{arikan2011systematic}
Erdal Arikan.
\newblock Systematic polar coding.
\newblock {\em IEEE communications letters}, 15(8):860--862, 2011.

\bibitem{bennett1992experimental}
Charles~H Bennett, Fran{\c{c}}ois Bessette, Gilles Brassard, Louis Salvail, and
  John Smolin.
\newblock Experimental quantum cryptography.
\newblock {\em Journal of cryptology}, 5(1):3--28, 1992.

\bibitem{bennett2014quantum}
Charles~H Bennett and Gilles Brassard.
\newblock Quantum cryptography: public key distribution and coin tossing.
\newblock {\em Theor. Comput. Sci.}, 560(12):7--11, 2014.

\bibitem{bennett1995generalized}
Charles~H Bennett, Gilles Brassard, Claude Cr{\'e}peau, and Ueli~M Maurer.
\newblock Generalized privacy amplification.
\newblock {\em IEEE Transactions on Information Theory}, 41(6):1915--1923,
  1995.

\bibitem{brassard1993secret}
Gilles Brassard and Louis Salvail.
\newblock Secret-key reconciliation by public discussion.
\newblock In {\em Workshop on the Theory and Application of of Cryptographic
  Techniques}, pages 410--423. Springer, 1993.

\bibitem{buttler2003fast}
William~T Buttler, Steven~K Lamoreaux, Justin~R Torgerson, GH~Nickel,
  CH~Donahue, and Charles~G Peterson.
\newblock Fast, efficient error reconciliation for quantum cryptography.
\newblock {\em Physical Review A}, 67(5):052303, 2003.

\bibitem{che2017physical}
Zhen Che, Junbin Fang, Zoe~Lin Jiang, Xiaolong Yu, Guikai Xi, and Zhe Chen.
\newblock A physical-layer secure coding schcme for visible light communication
  based on polar codes.
\newblock In {\em 2017 Conference on Lasers and Electro-Optics Pacific Rim
  (CLEO-PR)}, pages 1--2. IEEE, 2017.

\bibitem{chen2018secret}
Bin Chen and Frans~MJ Willems.
\newblock Secret key generation over biased physical unclonable functions with
  polar codes.
\newblock {\em IEEE Internet of Things Journal}, 6(1):435--445, 2018.

\bibitem{diamanti2016practical}
Eleni Diamanti, Hoi-Kwong Lo, Bing Qi, and Zhiliang Yuan.
\newblock Practical challenges in quantum key distribution.
\newblock {\em npj Quantum Information}, 2(1):1--12, 2016.

\bibitem{elkouss2009efficient}
David Elkouss, Anthony Leverrier, Romain All{\'e}aume, and Joseph~J Boutros.
\newblock Efficient reconciliation protocol for discrete-variable quantum key
  distribution.
\newblock In {\em 2009 IEEE International Symposium on Information Theory},
  pages 1879--1883. IEEE, 2009.

\bibitem{elkouss2010information}
David Elkouss, Jesus Martinez-Mateo, and Vicente Martin.
\newblock Information reconciliation for quantum key distribution.
\newblock {\em arXiv preprint arXiv:1007.1616}, 2010.

\bibitem{fung2010practical}
Chi-Hang~Fred Fung, Xiongfeng Ma, and HF~Chau.
\newblock Practical issues in quantum-key-distribution postprocessing.
\newblock {\em Physical Review A}, 81(1):012318, 2010.

\bibitem{gisin2002quantum}
Nicolas Gisin, Gr{\'e}goire Ribordy, Wolfgang Tittel, and Hugo Zbinden.
\newblock Quantum cryptography.
\newblock {\em Reviews of modern physics}, 74(1):145, 2002.

\bibitem{gottesman2004security}
Daniel Gottesman, H-K Lo, Norbert Lutkenhaus, and John Preskill.
\newblock Security of quantum key distribution with imperfect devices.
\newblock In {\em International Symposium onInformation Theory, 2004. ISIT
  2004. Proceedings.}, page 136. IEEE, 2004.

\bibitem{gottesman2003proof}
Daniel Gottesman and Hoi-Kwong Lo.
\newblock Proof of security of quantum key distribution with two-way classical
  communications.
\newblock {\em IEEE Transactions on Information Theory}, 49(2):457--475, 2003.

\bibitem{Hayashi2011Exponential}
M.~Hayashi.
\newblock Exponential decreasing rate of leaked information in universal random
  privacy amplification.
\newblock {\em IEEE Transactions on Information Theory}, 57(6):3989--4001,
  2011.

\bibitem{jouguet2012high}
Paul Jouguet and Sebastien Kunz-Jacques.
\newblock High performance error correction for quantum key distribution using
  polar codes.
\newblock {\em arXiv preprint arXiv:1204.5882}, 2012.

\bibitem{kim2017reconciliation}
Yongseen Kim, Changho Suh, and June-Koo~Kevin Rhee.
\newblock Reconciliation with polar codes constructed using gaussian
  approximation for long-distance continuous-variable quantum key distribution.
\newblock In {\em 2017 International Conference on Information and
  Communication Technology Convergence (ICTC)}, pages 301--306. IEEE, 2017.

\bibitem{koashi2009simple}
M~Koashi.
\newblock Simple security proof of quantum key distribution based on
  complementarity.
\newblock {\em New Journal of Physics}, 11(4):045018, 2009.

\bibitem{lee2018improved}
Sunghoon Lee, Jooyoun Park, and Jun Heo.
\newblock Improved reconciliation with polar codes in quantum key distribution.
\newblock {\em arXiv preprint arXiv:1805.05046}, 2018.

\bibitem{leverrier2009unconditional}
Anthony Leverrier and Philippe Grangier.
\newblock Unconditional security proof of long-distance continuous-variable
  quantum key distribution with discrete modulation.
\newblock {\em Physical review letters}, 102(18):180504, 2009.

\bibitem{Lijing2019}
Jin Li, Lin Jiang, Xucheng Lin, and Junbin Fang.
\newblock Polar codes-based one-step post-processing for quantum key
  distribution.
\newblock {\em Journal of South China Normal University (Natural Science
  Edition)}, 51(2):1--6, 2019.

\bibitem{LiHigh}
Qiong Li, Bing-Ze Yan, Hao-Kun Mao, Xiao-Feng Xue, Qi~Han, and Hong Guo.
\newblock High-speed and adaptive fpga-based privacy amplification in quantum
  key distribution.
\newblock {\em IEEE Access}, 7:21482--21490.

\bibitem{lo2001proof}
Hoi-Kwong Lo.
\newblock Proof of unconditional security of six-state quantum key distribution
  scheme.
\newblock {\em arXiv preprint quant-ph/0102138}, 2001.

\bibitem{lo2001simple}
Hoi-Kwong Lo.
\newblock A simple proof of the unconditional security of quantum key
  distribution.
\newblock {\em Journal of Physics A: Mathematical and General}, 34(35):6957,
  2001.

\bibitem{lo1999unconditional}
Hoi-Kwong Lo and Hoi~Fung Chau.
\newblock Unconditional security of quantum key distribution over arbitrarily
  long distances.
\newblock {\em science}, 283(5410):2050--2056, 1999.

\bibitem{mahdavifar2011}
Hessam Mahdavifar and Alexander Vardy.
\newblock Achieving the secrecy capacity of wiretap channels using polar codes.
\newblock {\em IEEE Transactions on Information Theory}, 57(10):6428--6443,
  2011.

\bibitem{mao2019high}
Haokun Mao, Qiong Li, Qi~Han, and Hong Guo.
\newblock High-throughput and low-cost ldpc reconciliation for quantum key
  distribution.
\newblock {\em Quantum Information Processing}, 18(7):232, 2019.

\bibitem{molotkov2001simple}
Sergey~N Molotkov and Sergey~S Nazin.
\newblock A simple proof of unconditional security of relativistic quantum
  cryptography.
\newblock {\em Journal of Experimental and Theoretical Physics},
  92(5):871--878, 2001.

\bibitem{mori2010properties}
Ryuhei Mori.
\newblock Properties and construction of polar codes.
\newblock {\em arXiv preprint arXiv:1002.3521}, 2010.

\bibitem{mori2009performance}
Ryuhei Mori and Toshiyuki Tanaka.
\newblock Performance and construction of polar codes on symmetric binary-input
  memoryless channels.
\newblock In {\em 2009 IEEE International Symposium on Information Theory},
  pages 1496--1500. IEEE, 2009.

\bibitem{nakassis2017polar}
Anastase Nakassis.
\newblock Polar codes for quantum key distribution.
\newblock {\em Journal of Research of the National Institute of Standards and
  Technology}, 122, 2017.

\bibitem{pearson2004high}
David Pearson.
\newblock High-speed qkd reconciliation using forward error correction.
\newblock In {\em AIP Conference Proceedings}, volume 734, pages 299--302.
  American Institute of Physics, 2004.

\bibitem{renes2013efficient}
Joseph~M Renes, Renato Renner, and David Sutter.
\newblock Efficient one-way secret-key agreement and private channel coding via
  polarization.
\newblock In {\em International Conference on the Theory and Application of
  Cryptology and Information Security}, pages 194--213. Springer, 2013.

\bibitem{shor2000simple}
Peter~W Shor and John Preskill.
\newblock Simple proof of security of the bb84 quantum key distribution
  protocol.
\newblock {\em Physical review letters}, 85(2):441, 2000.

\bibitem{tal2013construct}
Ido Tal and Alexander Vardy.
\newblock How to construct polar codes.
\newblock {\em IEEE Transactions on Information Theory}, 59(10):6562--6582,
  2013.

\bibitem{walenta2014fast}
Nino Walenta, Andreas Burg, Dario Caselunghe, Jeremy Constantin, Nicolas Gisin,
  Olivier Guinnard, Rapha{\"e}l Houlmann, Pascal Junod, Boris Korzh, Natalia
  Kulesza, et~al.
\newblock A fast and versatile quantum key distribution system with hardware
  key distillation and wavelength multiplexing.
\newblock {\em New Journal of Physics}, 16(1):013047, 2014.

\bibitem{wyner1975wire}
Aaron~D Wyner.
\newblock The wire-tap channel.
\newblock {\em Bell system technical journal}, 54(8):1355--1387, 1975.

\bibitem{yan2018improved}
Shiling Yan, Jindong Wang, Junbin Fang, Lin Jiang, and Xuan Wang.
\newblock An improved polar codes-based key reconciliation for practical
  quantum key distribution.
\newblock {\em Chinese Journal of Electronics}, 27(2):250--255, 2018.

\bibitem{yi2019efficient}
Zhengzhong Yi, Junbin Fang, Puxi Lin, Xiaojun Wen, Zoe~Lin Jiang, and Xuan
  Wang.
\newblock Efficient quantum key distribution protocol based on
  classical--quantum polarized channels.
\newblock {\em Quantum Information Processing}, 18(12):356, 2019.

\bibitem{yuan201810}
Zhiliang Yuan, Alan Plews, Ririka Takahashi, Kazuaki Doi, Winci Tam, Andrew
  Sharpe, Alexander Dixon, Evan Lavelle, James Dynes, Akira Murakami, et~al.
\newblock 10-mb/s quantum key distribution.
\newblock {\em Journal of Lightwave Technology}, 36(16):3427--3433, 2018.

\bibitem{zhang2012real}
Hong-Fei Zhang, Jian Wang, Ke~Cui, Chun-Li Luo, Sheng-Zhao Lin, Lei Zhou, Hao
  Liang, Teng-Yun Chen, Kai Chen, and Jian-Wei Pan.
\newblock A real-time qkd system based on fpga.
\newblock {\em Journal of Lightwave Technology}, 30(20):3226--3234, 2012.

\end{thebibliography}


\end{document}